# From Extraterrestrial Microbes to Alien Intelligence:

# Rebalancing Astronomical Research Priorities

Omer Eldadi[1], Gershon Tenenbaum[1] and Abraham Loeb[2]

1. B. Ivcher School of Psychology, Reichman University, Herzliya, Israel
2. Department of Astronomy, Harvard University, Cambridge, MA, USA

**Corresponding Address**:

Omer Eldadi

B.Ivcher School of Psychology

Reichman University

The University 8, Herzliya, Israel

Email: Omereldadi@gmail.com

**Abstract**

We examine the funding disparity in astronomical research priorities: the Habitable Worlds Observatory is planned to receive over $10 billion over the next two decades whereas extraterrestrial intelligence research receives nearly zero federal funding. This imbalance is in contrast to both scientific value and public interest, as 65% of Americans and 58.2% of surveyed astrobiologists believe extraterrestrial intelligence exists. Empirical psychological research demonstrates that humanity possesses greater resilience toward extraterrestrial contact than historically recognized. Contemporary studies reveal adaptive responses rather than mass panic, conflicting with the rationale for excluding extraterrestrial intelligence research from federal funding since 1993. The response to the recent interstellar object 3I/ATLAS exemplifies consequences of this underinvestment: despite discovery forecasts of a new interstellar object every few months for the coming decade, no funded missions exist to intercept or closely study these visitors from outside the Solar System. We propose establishing a comprehensive research program to explore both biosignatures and technosignatures on interstellar objects. This program would address profound public interest while advancing detection capabilities and enabling potentially transformative discoveries in the search for extraterrestrial life. The systematic exclusion of extraterrestrial intelligence research represents institutional bias rather than scientific limitation, requiring immediate reconsideration of funding priorities.

## Introduction

Why does modern science, despite claims of serving societal interests, ignore questions that captivate billions of people? The search for extraterrestrial intelligence offers a stark example for this puzzling question: while multiple surveys consistently show that 58.2%-65% of the participants believe extraterrestrial intelligence exists (Pew Research Center, 2021; Vickers et al., 2025), searching for past or present technosignatures beyond Earth falls within the capabilities of current and planned planetary science missions (Astro2020, 2023; Haqq-Misra et al., 2022). The National Academies' 2020 Decadal Survey of Astronomy and Astrophysics (Astro2020, 2023) exemplifies this disconnect, relegating the search for extraterrestrial intelligence entirely to private funding (Haqq-Misra et al., 2022; Wright & Oman-Reagan, 2018) while federally funded projects are overwhelmingly focused on detecting microbial biosignatures (Astro2020, 2023).

The magnitude of this disparity is striking: more than ten billion dollars are planned to be allocated over the next two decades to the "Habitable Worlds Observatory" (HWO; Astro2020, 2023), a space telescope that will search for spectral fingerprints of molecules indicative of microbes in the atmospheres of 25 carefully selected exoplanets (Mamajek & Stapelfeldt, 2024). This mission explicitly focuses on biosignatures (e.g., oxygen, methane, water vapor, and other indicators of biological processes), while excluding any capability for extraterrestrial intelligence detection. Meanwhile, interstellar objects, which can potentially harbor evidence of extraterrestrial intelligence (Loeb 2022, Ezell & Loeb, 2023), receive only a brief mention in the decadal survey despite the document's acknowledgment of their scientific significance.

The survey's own language reveals this contradiction: "One of the most exciting astronomical discoveries of the past few years is two interstellar interlopers, the asteroid

'Oumuamua and the comet 2I/Borisov, that originated around another star and passed through our solar system. Large-scale surveys such as the *Rubin Observatory Legacy Survey of Space and Time* will discover many more such objects and will provide an increased understanding of context and impact for these interstellar visitors" (Astro2020, 2023, p. 50). Yet, this proclaimed excitement contrasts sharply with the absence of dedicated federal funding for intercepting or closely studying these visitors from other star systems.

The new interstellar object 3I/ATLAS crystallizes the consequences of this misalignment. When it reaches perihelion on October 29th, 2025, 3I/ATLAS will pass on the opposite side of the Sun relative to Earth, making terrestrial observations of its brightest phase impossible (Hibberd, Crowl & Loeb, 2025; Loeb, 2025a). As this visitor passes by at up to ~68 km/s in the opposite direction to Earth's motion around the Sun, there will not be sufficient opportunity to study it due to decades of underinvestment in the necessary interception and close observation capabilities. As Siraj et al. (2023) revealed, intercepting interstellar objects along their path through the Solar System requires pre-positioned spacecraft, which are not currently available. In this article, we suggest that the systematic exclusion of the search for extraterrestrial artifacts from mainstream research, despite the technological feasibility of this search (Hein et al., 2022) and overwhelming public interest in it, reveals an institutional bias against novel research paradigms, indicating it is time for astronomers to embrace a bolder scientific program.

## Theoretical Background

### Extraordinary Claims Require Extraordinary Evidence and Extraordinary Funding

The systematic underfunding of technosignature research stems from the misapplication of Carl Sagan's dictum "Extraordinary Claims Require Extraordinary Evidence", articulated in 1979 (Sagan, 1979). While Sagan intended this as a call for methodological rigor, it has evolved into a gatekeeping mechanism that reflexively dismisses extraterrestrial intelligence research as inherently "extraordinary". Yet, given an estimated $10^{22}$ habitable planets in the observable volume of the Universe (Loeb, 2016), the assumption that Earth hosts the only technological civilization represents the truly "extraordinary claim".

Additionally, this conceptual framing fundamentally misrepresents the nature of evidence. Evidence becomes "extraordinary" when obtained without commensurate investment. Discovering extraterrestrial intelligence technosignature will probably require decades and billions of dollars in funding, making the discovery ordinary rather than "extraordinary". Therefore, we propose that Sagan's principle requires amendment: "scientific hypotheses require evidence to test them and sufficient funding to collect that evidence". By classifying extraterrestrial intelligence research as extraordinary while withholding appropriate resources, the scientific community creates a self-fulfilling prophecy, ensuring that the "extraordinary evidence" remains out of reach.

This misapplication of Sagan's principle extends beyond extraterrestrial intelligence research, threatening scientific innovation more broadly. As Deming conclude, "Ideas, theories, or observations that are merely novel are not "extraordinary", nor do they require an "extraordinary" amount of evidence for corroboration. Science does not contemplate two types of evidence. The misuse of "Extraordinary Claims Require Extraordinary Evidence" to suppress innovation and maintain orthodoxy should be avoided as it must inevitably retard the progress of science in establishing comprehensive and systematic bodies of reliable knowledge" (Deming, 2016, p.11).

**The Conservative Paradigm and Its Consequences**

NASA's Search for Extraterrestrial Intelligence (SETI) program exemplifies how social stigma can override scientific merit in funding decisions. The program's vulnerability stemmed not from scientific inadequacy but from the "giggle factor"—the tendency for SETI to provoke dismissive laughter rather than serious consideration (Wright & Oman-Reagan, 2018). The social stigma manifested through congressional ridicule, with Senator Richard Bryan successfully championing the program's termination in 1993, declaring his amendment would "end the Great Martian Chase at the taxpayers' expense" (Garber, 1999), effectively excluding an entire domain of scientific inquiry from public funding.

The quantitative consequences of this decision reveal funding asymmetries in astronomy research priorities. While NASA has invested substantially in astrobiology research, with annual budgets reaching approximately $65 million in recent years (NASA Astrobiology Program, n.d.), extraterrestrial intelligence research has received minimal federal funding since 1993. The HWO alone will command more than $10 billion over the next two decades (Astro2020, 2023), while extraterrestrial intelligence research survives through private philanthropy, most notably Breakthrough Listen (Worden et al., 2017) and the donated endowment of the SETI Institute. This creates a funding disparity between the search for microbial biosignatures and extraterrestrial artifacts.

The channeling towards private funding has created a self-reinforcing marginalization cycle. NASA's 2015 Astrobiology Strategy explicitly states: "While traditional SETI is not part of astrobiology, and is currently well-funded by private sources, it is reasonable for astrobiology to maintain strong ties to the SETI community" (Hays, 2015, p. 150). Such a circular reasoning, excluding SETI because private funding exists, while private funding exists only due to exclusion from federal funding, extends to practical consequences. When organizations like Messaging to Extraterrestrial Intelligence (METI) International conduct active transmission efforts, critics raise legitimate concerns about attracting hostile civilizations—issues that would benefit from coordinated government oversight rather than emerging from an unregulated private sector (Gertz, 2016). Yet, these funding disparities cannot be fully explained by practical concerns or historical contingencies alone; they reflect deeper epistemological assumptions about what constitutes legitimate scientific inquiry.

**The Social Construction of Scientific Priorities**

The systematic underfunding of extraterrestrial intelligence research cannot be explained through purely epistemic considerations. As Pellegrini (2024) reveals in the continental drift controversy, the social acceptance of scientific theories, and by extension, research priorities, depends not on intrinsic rational merit but on alignment with prevailing institutional "styles of thought". This conceptual framework illuminates the reasons biosignature research receives billions of dollars whereas extraterrestrial intelligence research receives negligible federal funding, despite their comparable scientific merit (Lingam & Loeb, 2019).

Scientific funding decisions reflect "cultural conditioning of knowledge", where certain research questions become acceptable or unacceptable based on institutional dynamics rather than objective evaluation (Pellegrini, 2024). The astronomical community's embrace of microbial biosignatures while rejecting extraterrestrial intelligence research exemplifies this phenomenon. Both searches rest on similar probabilistic arguments about life's emergence, yet only one fits comfortably within established paradigms that separate "serious" astrobiology from "fringe" extraterrestrial intelligence search.

Such an institutional bias exemplifies what Barnes and Bloor (1982) critique as the rationalist separation of social and epistemic factors in explaining scientific development. Funding agencies justify their preferences by claiming that biosignature research is more "rigorous" or "fundamental" or "likely to succeed", yet these assessments reflect pre-existing cultural assumptions about what constitutes legitimate science. The result is a self-reinforcing cycle: extraterrestrial intelligence research lacks credibility because it lacks funding, and it lacks funding because it lacks institutional credibility.

The funding priorities reveal inconsistent application of scientific standards. Federal agencies have invested vast amount of funds in dark matter detection experiments (Billard et al., 2022), despite persistent null results. Yet the existence of technological civilizations has at least one confirmed example—our own—while dark matter remains entirely theoretical and could potentially be replaced by a modified behavior of gravity at low accelerations. This disparity suggests that funding decisions reflect institutional comfort with established research paradigms rather than objective assessment of scientific merit or societal interest. Addressing this imbalance requires recognizing that institutional dynamics, not scientific limitations, have shaped current funding priorities, and that advancing transformative scientific questions demands both methodological rigor and willingness to challenge these established paradigms. While institutional dynamics have systematically marginalized extraterrestrial intelligence research, mounting evidence suggests this conservatism diverges sharply from widespread public fascination with extraterrestrial intelligence and readiness for potential discoveries.

**Public Interest in Extraterrestrial Intelligent Life**

The search for extraterrestrial intelligence has captivated public imagination for decades, with contemporary surveys revealing substantial belief across diverse populations. While some studies measure belief in extraterrestrial life broadly, with 90% of Swedish students (Persson et al., 2019) and 92% of Peruvian university students (Chon-Torres et al., 2020) expressing such beliefs, the most relevant data comes from surveys specifically addressing extraterrestrial intelligent life. When asked directly about extraterrestrial intelligent life, 65% of Americans express belief in its existence (Pew Research Center, 2021). Additionally, Vickers et al. (2025), who found that while 86.6% of astrobiologists believe basic extraterrestrial life likely exists, confidence drops to 67.4% for complex life and 58.2% for intelligent life. That most astrobiologists experts in the field consider extraterrestrial intelligence likely underscores the scientific legitimacy of the search for extraterrestrial intelligence. Such an expert consensus challenges the marginalization of intelligence-focused research programs and supports the reallocation of resources toward extraterrestrial technological signatures right now.

The current literature reveals a critical gap that strengthens the case to reinforce technosignatures research. While existing surveys reveal substantial public belief in extraterrestrial intelligence, no comprehensive studies examine public preferences for the split in funding allocations between biosignatures and technosignatures search programs. The absence of granular data represents an untapped opportunity to reveal public support for technosignatures research. Given that 58.2-65% of populations (including astrobiologist experts) specifically endorse the existence of extraterrestrial intelligent life when directly asked, investigating their funding preferences could reveal substantial backing to explore the search for extraterrestrial intelligence, currently relegated to private philanthropy at the periphery of astrobiology.

**The Rationalist Fallacy and Public Dismissal**

The disconnect between public interest and funding priorities may reflect what Pellegrini (2024) identifies as a rationalist approach that "tends to judge rather than understand" (p. 921).

Drawing from Pellegrini's analysis of rationalist approaches, which he notes share similarities with what he identifies as the deficit model (p. 915), labeling those in disagreement as "wrong", we can analyze how funding decisions may dismiss public interest in extraterrestrial intelligence as scientific illiteracy rather than recognizing it as reflecting legitimate cultural values and contexts.

This fallacy operates through what science communication scholars term the "deficit model", which presumes that public enthusiasm for extraterrestrial intelligence search will diminish with proper scientific education. However, our data showing that 58.2% of astrobiologists believe in extraterrestrial intelligence contradicts this assumption. The public's position aligns with expert opinion, suggesting their mutual interests reflects informed judgment rather than naive fantasy.

Pellegrini's distinction between "judging" and "understanding" social phenomena proves crucial here. Funding agencies judge public interest as irrational while failing to understand the sophisticated reasoning underlying this interest. The public intuitively grasps what Wright et al. (2022) demonstrate quantitatively: technological signatures may offer detection advantages spanning 20 orders of magnitude compared to biosignatures. The public preference for technosignatures research reflects sound probabilistic thinking, not scientific ignorance.

**The Scientific Case for Extraterrestrial Intelligence Search Prioritization**

The scientific legitimacy of the scientific search for extraterrestrial intelligence, supported by majority expert opinion (Vickers et al., 2025), stands in stark contrast to its institutional marginalization. Five decades ago, Sagan and Drake (1975) presciently argued that "There can be little doubt that civilizations more advanced than the Earth's exist elsewhere in the universe. The probabilities involved in locating one of them call for a substantial effort" (p. 80).

This early recognition, combined with sustained public interest and expert consensus, makes current funding allocations particularly problematic.

Wright et al. (2022) present a paradigm-shifting analysis demonstrating that it is also plausible that N(tech) ≫ N(bio), based on technology's unique properties. First, technology possesses spreading capabilities through interstellar colonization and self-replicating probes, with potential abundance increases up to $10^{10}$ in extreme scenarios. Second, certain technosignatures exhibit remarkable longevity—solar collectors on airless worlds could persist for hundreds of millions of years, far exceeding biological constraints. Third, detection advantages span over 20 orders of magnitude on the Kardashev scale, from local emissions to stellar-scale energy manipulation. Finally, specific technosignatures like narrowband radio signals offer unambiguous artificiality, eliminating the false-positive challenges inherent to biosignatures interpretation.

These compelling theoretical advantages raise a critical question: If researchers of extraterrestrial intelligence technosignatures contemplate superior detection prospects that may yield results sooner than biosignature programs, is humanity prepared for such a discovery? The potential for encountering unambiguous evidence of extraterrestrial civilizations, whether through narrowband signals, megastructures, or other technosignatures, necessitates examining not only our scientific readiness but also our societal capacity to process such paradigm-shifting information. While historical concerns about mass panic have influenced policy decisions and funding priorities, contemporary research reveals a remarkably different picture of human psychological resilience.

**Public Resilience and Societal Readiness for Contact**

Historical assessments of public readiness for contact with extraterrestrial intelligence have profoundly influenced current policy and scientific discourse. The Robertson Panel's 1953 warning about potential mass hysteria (Durant, 1953) established a narrative of inevitable public panic. This pessimistic viewpoint persists in a curious form: while 25% of Americans anticipate widespread fear following contact, this concern reflects their perception of how others would react rather than their own anticipated response (Harrison, 2011). The same study also found that over 60% of respondents considered themselves personally invulnerable or fully capable of adapting to extraterrestrial intelligence revelations. Such a disconnects between perceived societal fragility and personal resilience exemplifies the "third-person effect" (Davison, 1983), where individuals systematically overestimate others' vulnerability while expressing confidence in their own psychological stability. These findings challenge the foundational assumption of mass panic that has long justified limiting public engagement with extraterrestrial intelligence research.

Direct empirical research on reactions to intelligent extraterrestrial contact remains remarkably limited. Vakoch and Lee (2000) developed psychometrically validated scales to assess multidimensional beliefs about extraterrestrial intelligence following a hypothetical message receipt. Their cross-cultural study of American and Chinese undergraduates revealed that respondents could simultaneously view extraterrestrial intelligence as potentially benevolent and malevolent, particularly among Chinese participants, where these beliefs showed near-zero correlation. These findings challenge assumptions that public reactions can be uniformly negative or catastrophic.

While Kwon et al. (2018) primarily examined reactions to discoveries of microbial life, they highlighted crucial observational evidence. Citing survey data (Main, 2016), they noted that despite the majority belief in extraterrestrial life among Americans, British, and Germans, with substantial percentages believing Earth has already been visited, "in none of these societies have we seen an utter breakdown in social order or panic as a result of these widespread beliefs" (Kwon et al., 2018, p.8). This extended observational period provides compelling evidence against catastrophic reaction hypotheses.

Theoretical frameworks from social psychology offer additional insights. Social Identity Theory (SIT; Tajfel & Turner, 1979) suggests that contact with out-groups (in our case extraterrestrial intelligence), could catalyze formation of a superordinate "Earthlings" identity, transcending traditional terrestrial divisions. This phenomenon mirrors documented cases where external threats unite previously antagonistic groups (Gaertner & Dovidio, 2000; Sherif et al., 1961). Rather than fragmenting humanity, confirmation of extraterrestrial intelligence, or even the systematic search itself, could catalyze unprecedented global unity.

Lingam et al. (2023) developed rigorous statistical methods for evaluating potential observable indicators of alien technology. The publication of such methodological frameworks in peer-reviewed venues reflects a broader recognition that this research advances scientific methodology and inspires public engagement with science, regardless of whether contact occurs.

This convergence of limited but consistent empirical findings (Vakoch & Lee 2000), robust theoretical frameworks (Gaertner & Dovidio, 2000; Sherif et al., 1961; Tajfel & Turner, 1979), and extensive observational evidence (Harrison, 2011; Kwon et al., 2018), indicate that humanity possesses sufficient psychological and social structures to integrate profound discoveries constructively. Decisions regarding research funding must reflect evidence-based assessments rather than perpetuating unfounded assumptions about public panic. Yet, while humanity reveals psychological readiness for contact, systematic underinvestment in extraterrestrial intelligence search continues to result in missed opportunities. Nowhere is this more evident than in our inability to study interstellar objects—potential carriers of both natural and artificial information from other star systems.

**The Gift of Interstellar Objects: The 3I/ATLAS Case**

The approaching 3I/ATLAS exemplifies the consequences of systematic underinvestment in interstellar objects research. The visitor from another star system will pass at ~60 km/s on October 29, 2025 (Seligman et al., 2025; Loeb, 2025a,b), with no existing capability to intercept or closely examine it. This represents the third such missed opportunity for a flyby or rendezvous mission within a decade, following 1I/'Oumuamua (2017) and 2I/Borisov (2019).

The scientific significance of interstellar objects research extends across multiple disciplines, offering unprecedented opportunities to study material from beyond our solar system. The detection of 'Oumuamua and subsequent interstellar visitors has opened new avenues for investigating fundamental questions about planetary system formation and galactic processes. Hein et al. (2022) provides a comprehensive synthesis of these multidisciplinary research opportunities:

"These interstellar objects provide a previously inaccessible opportunity to directly sample physical material from other stellar systems much sooner than otherwise. By analyzing these interstellar interlopers, we can acquire significant data and deduce information about their planetary system of origin (Feng and Jones, 2018; Portegies Zwart et al., 2018; Moro-Martín, 2018; Jackson et al., 2018), planetary formation (Trilling et al., 2017; Raymond et al., 2018; Rice and Laughlin, 2019), galactic evolution, and possibly molecular biosignatures (Lingam and Loeb, 2018) or even clues about panspermia (Ginsburg et al., 2018)". (p. 403)

This comprehensive research agenda has gained particular urgency as detection capabilities improve. Compositional analysis through isotopic ratios can reveal nucleosynthetic histories that vary across galactic regions, potentially pinpointing these objects' origins (Nittler & Gaidos, 2012). Beyond planetary science applications, interstellar objects also present opportunities to search for potential biosignatures or extraterrestrial intelligence technosignatures, including the possibility of artificial origins (Bialy & Loeb, 2018; Loeb, 2022; Loeb, 2025c). Direct sampling or spectroscopic analysis of interstellar material during close encounters can provide unprecedented data to test hypotheses about the distribution of life throughout the galaxy.

Despite projections that the Vera Rubin Observatory will detect 1-10 interstellar objects annually (Dorsey et al. 2025; Siraj et al., 2023), no funded missions currently exist to study or intercept these visitors. The development of intercept capabilities following 'Oumuamua's discovery could have enabled transformative science comparable to or exceeding that of current flagship missions (Hein et al., 2019; Siraj et al., 2023).

**Institutional Dynamics and Path Dependence**

The current funding landscape exemplifies what organizational theorists term "path dependence"—where historical decisions constrain future possibilities regardless of changing circumstances. The cancellation of SETI's federal funding in 1993 created institutional precedents that persist despite dramatic improvements in detection capabilities and theoretical understanding (Garber, 1999; Haqq-Misra et al., 2022).

Path dependence operates through multiple mechanisms. First, the absence of extraterrestrial intelligence technosignature experts in funding review panels ensures proposals face evaluation by scientists socialized to view such research skeptically and prefer that funding to entrenched astrobiology programs in which they are engaged will not be shared with competing research programs. Second, early-career researchers avoid technosignature topics to maintain funding viability, creating generational reproduction of bias. Third, the private funding, rather than demonstrating the field's vitality, provides agencies an excuse to maintain exclusion—a circular logic Pellegrini (2024) identifies in other scientific controversies.

Breaking this cycle requires recognizing that current funding patterns reflect contingent historical choices, not inevitable scientific logic. The continental drift controversy discussed by Pellegrini (2024), revealed how entrenched paradigms can persist for decades despite mounting evidence. Similarly, technosignature research may require institutional disruption, perhaps through congressional mandate or international competition, to overcome accumulated structural barriers.

## Discussion

The evidence presented here reveals a systematic misalignment between astronomical funding priorities and converging lines of scientific, technological, and societal readiness. The funding disparity between biosignature and technosignature research is unreasonably extreme (Astro2020, 2023; Haqq-Misra et al., 2022). This asymmetry persists despite 58.2% of astrobiologists acknowledging the likelihood of extraterrestrial intelligence (Vickers et al., 2025) and theoretical analyses suggesting extraterrestrial intelligence technosignatures may offer superior detection prospects due to their longevity, spreading capability, and unambiguous artificiality (Wright et al., 2022).

The new 3I/ATLAS (Loeb, 2025; Seligman et al., 2025) encounter crystallizes the consequences of this underinvestment. As the design of flyby or rendezvous missions with interstellar objects through the inner solar system was underfunded over the past decade, this neglect resulted in lost data about other stellar systems—precisely the "previously inaccessible opportunity to directly sample physical material from other stellar systems" that Hein et al. (2022, p. 403) identified. With projections indicating the Vera C. Rubin Observatory will detect 1-10 such objects annually (Dorsey et al. 2025; Siraj et al., 2023), the absence of intercept capabilities transforms predictable opportunities into systematic losses in our scientific knowledge.

Direct empirical evidence contradicts the historical assumptions on public reactions to extraterrestrial contact scenarios and reveals resilience rather than panic (Vakoch & Lee, 2000), while observational data shows that widespread belief in extraterrestrial visitation has produced no social breakdown (Kwon et al., 2018). The "third-person effect" identified by Harrison (2011), where individuals expect others to panic while expressing personal confidence, explains how unfounded assumptions about public fragility have persisted despite contradictory evidence.

The scientific case for rebalancing priorities is compelling. As Sagan and Drake (1975) presciently argued, "There can be little doubt that civilizations more advanced than the earth's exist elsewhere in the universe. The probabilities involved in locating one of them call for a substantial effort" (p. 80). Five decades later, improved detection capabilities and validated public resilience makes continued marginalization of extraterrestrial intelligence research increasingly untenable. The technological innovations required for interstellar object interception would yield benefits across multiple sectors, as revealed by historical precedents in ambitious space programs (Hein et al., 2019). Additionally, the asymmetric potential of extraterrestrial intelligence discoveries justifies proportionate funding; while biosignatures may accumulate incremental evidence for microbial life, single confirmed evidence for extraterrestrial intelligence will instantaneously transform humanity's cosmic perspective.

**Moving Forward: From Documentation to Transformation**

This analysis reveals that the funding crisis of extraterrestrial intelligence research stems not from scientific deficiencies but from institutional dynamics that privilege established research paradigms over public interests and emerging possibilities. Following Pellegrini's (2024) conceptual framework, we must move beyond documenting inequality to understanding and transforming the social processes that perpetuating it. Concrete steps toward rebalancing priorities include: (1) democratizing review processes, include public representatives and social scientists on astronomy funding panels to counterbalance disciplinary insularity, (2) recognizing extraterrestrial intelligence research as mainstream astronomy, (3) including extraterrestrial intelligence research within the core mission of the next Decadal Survey of Astronomy & Astrophysics, not relegated to appendices, mandating by U.S. Congress of a minimum percentages for high-risk, high-reward research aligned with public interests, and (4) reframing the narrative of the intellectual discourse to acknowledge that assuming Earth's unique technological status constitutes the truly extraordinary claim given the vast cosmic scales.

The unfulfilled interception mission with 3I/ATLAS symbolizes the perpetuate institutional biases that led to missed opportunities instead of embracing bold scientific programs. The convergence of public interest, technological capability, and cosmic likelihood delivers an unprecedented opportunity to learn, if our institutions will possess the wisdom to seize it.

2020Durant, F. C. (1953). *Report of meetings of scientific advisory panel on unidentified flying objects convened by Office of Scientific Intelligence*. Washington, DC: Central Intelligence Agency.

Eldadi, O., Tenenbaum, G., & Loeb, A. (under review). Scientific paradigm resistance evidence from the Oumuamua debate and cross-disciplinary case. Submitted for publication.

Ezell, C., & Loeb, A. (2023). The inferred abundance of interstellar objects of technological origin. *Acta Astronautica*, 208, 124-129. https://doi.org/10.1016/j.actaastro.2023.03.030

Feng, F., & Jones, H. R. A. (2018). 'Oumuamua as a messenger from the Local Association. *The Astrophysical Journal Letters, 852*(2), L27. https://doi.org/10.3847/2041-8213/aaa404

Gaertner, S. L., & Dovidio, J. F. (2000). *Reducing intergroup bias: The common ingroup identity model*. Psychology Press.

Garber, S.J. (1999). Searching for good science – the cancellation of NASA's SETI program. *Journal of the British Interplanetary Society*, *52*(1), 3-12.

Gertz, J. (2016). *Reviewing METI: A critical analysis of the arguments* (arXiv:1605.05663v3). arXiv. https://doi.org/10.48550/arXiv.1605.05663

Ginsburg, I., Lingam, M., & Loeb, A. (2018). Galactic panspermia. *The Astrophysical Journal Letters, 868 (*L12). https://doi.org/10.3847/2041-8213/aaef2d

Harrison, A. A. (2011). The search for extraterrestrial intelligence: Astrosociology and cultural aspects. *Astropolitics, 9*(1), 63–83. https://doi.org/10.1080/14777622.2011.557619

Haqq-Misra, J., Ashtari, R., Benford, J., Carroll-Nellenback, J., Döbler, N. A., Farah, W., Fauchez, T. J., Gajjar, V., Grinspoon, D., Huggahalli, A., Kopparapu, R. K., Lazio, J., Profitiliotis, G., Sneed, E. L., Varghese, S. S., & Vidal, C. (2022). Opportunities for

...

```
```